# Slowing-down reduction and Possible Reversal Trend of Tropospheric NO2 over China during 2016 to 2019

Rui Li[1,2], Haixu Bo[1], Yu Wang[1,2,*]

1. School of Earth and Space Science, University of Science and Technology of China

2. Comparative Planetary Excellence Innovation Center, Chinese Academy of Sciences, Hefei, 230026, China

* Corresponding Author:   wangyu09@ustc.edu.cn

R. Li and H. Bo made equal contribution to this study

Atmospheric nitrogen dioxide ($NO_2$) over China at national level has been kept reducing since 2011 as seen from both satellite observations, ground-based measurements and bottom-up emission inventory (Liu et al., 2016; Irie et al., 2016; Krotkov et al., 2016; Foy et al., 2016; Liu et al 2017). These studies used data before 2015-2016. After 2016, however, a significant slowing-down of the reduction trend and/or even a reversal trend were found in numerous provinces, particularly in those with heavy $NO_2$ level, based on satellite observations. Error analysis on satellite data excluded cloud contamination, instrument anomalies from the main reasons of this change. Ground-based measurements show strong positive correlations with satellite observations and similar patterns of year-to-year changes of NO2 in 2018 and 2019 winter time. The temporal variations of Satellite NO2 over China are believed largely determined by surface emission from power plant and transportation. The reason for the recent change from emission perspective cannot be traced down since the national emission inventory was not updated since 2015. We therefore call on immediate attentions from both scientist community and policy makers to this phenomenon. Further efforts should be made to understand the reasons causing this change and to make associated air pollution controlling actions.

*Introduction*

Tropospheric nitrogen dioxide (NO2) and nitric oxide (NO) are important components of photochemical air pollutants which can cause serious harm to human's health (e.g. Mills et al., 2015; Luo et al., 2016). They also play determinate roles in the formation of secondary aerosols and ozone (Brewer et al., 1973; Seinfeld and Pandis, 2006), and in absorbing solar radiation in atmosphere (Solomon et al., 1999), and thus make direct and indirect impacts on global climate system.

China has been undergoing fast economic growth since 1990s with dramatic increase of consumption of fossil fuel from power plants, vehicles and facility construction. This has caused serious issue of air pollution over the country. To control the pollution, Chinese government enforced strict regulations and actions, such as putting desulfurization, denitrogenization devices and precipitators on power plants and vehicles. These efforts received immediate payback. For instance, investigations based on satellite observations, ground-based measurements and bottom-up inventory all agree that the national level of tropospheric $NO_2$ increased from 2005 (and even earlier, Richter et al., 2005) to 2011 but decreased from 2011 to 2015 (Liu et al., 2016; Irie et al., 2016; Krotkov et al., 2016; Foy et al., 2016;

Liu et al 2017). Quantitatively, the decreasing rate of NO2 column density derived from Ozone Monitoring Instrument (OMI) onboard the NASA Aura satellite was about 6-6.4% a year$^{-1}$ (Irie et al., 2016; Liu et al 2017). This is accompanied by the decline of surface NO2 emission at the pace of 4.2% a year$^{-1}$ (estimated from bottom-up inventory, Liu et al 2017). Base on this, the air pollution controlling goal of China's twelfth Five-Year Plan, i.e reducing national emissions of nitrogen oxides (NOx) by 10% by 2015 compared with 2010, was successfully achieved (Foy et al, 2016).

The thirteenth Five-Year Plan of China proposes to control total nitrogen dioxides emissions to be within 15.74 million tons by 2020, and a 15% decline from that in 2015 (Chinese Government Website). Few studies have been done to investigate the trend of $NO_2$ after 2015 using large scale satellite observations. In this study, we used both the NO2 column density derived from satellite (2005-2019) and NO2 concentration from ground sites (2015-2019) to study the trend of NO2 in the recent years.

*Results*

The column density of $NO_2$ in China mainly is determined by anthropogenic emission. Therefore, its spatial and temporal distribution is closely related to the status of the development and perturbations of economy. From the map of long-term mean NO2 derived from Aqua/OMI, column density of $NO_2$ is heavy (e.g. over 4 x $10^{15}$ mol cm$^{-2}$) in the south part of Northeast China (i.e. the provinces of HeilongJiang, Jiling and Liaoning), the Beijing-Tianjin-Hebei area, the Eastern China, etc. The time series of national averaged monthly mean NO2 (blue curve in Figure 1b) show significant seasonal cycle with peak value in January and valley value in June. The annual mean NO2 over China (the black curve, from 2005 to 2018) shows clear interannual variation of increasing from 2005 to 2011 and decreasing from 2011 to 2015/2016. The general pattern and this long-term trend of NO2 are consistent with the results from previous studies as mentioned in the introduction.

However, there is a change of trend in Figure 1b by looking at the time series after 2015-2016. First, the decreasing of NO2 from 2011 to 2015 clearly slowed down in the year 2016. Second, an increasing trend from 2016 to 2019 was shown in Figure 1b. To further understand the seasonal variation of such trend, we investigated the time series of seasonal mean $NO_2$ (i.e. winter of DJF, spring of MAM, summer of JJA and fall of SON) and found the increasing trend of annual mean NO2 was mainly contributed by the values in winter. The winter $NO_2$ (purple curve) significantly increased from 2017 to 2019 by ~25% per year. Now the mean $NO_2$ in 2019 DJF is very close to the levels at 2010DJF and 2014DJF. This introduces large uncertainty into the goal of China's thirteenth Five-Year Plan's pollution control measures, which plans a 15% reduction from 2015. For other seasons (green, red and orange curves for spring, summer and fall), although the mean values of NO2 is much smaller than that in winter, an increasing trend from 2016 to 2018 is very clear as well.

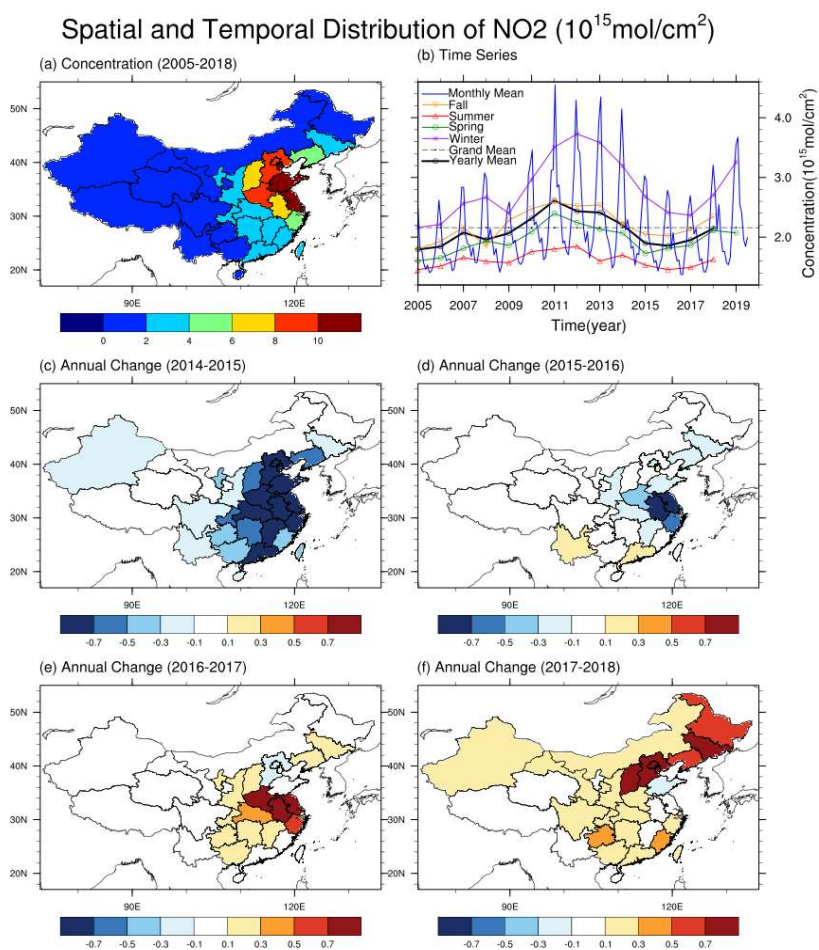

*Figure 1. The spatial and temporal distribution of NO$_2$ derived from Aura/OMI. a) Horizontal distribution of long-term mean (2005.1-2019.6) column NO$_2$ (unit:10$^{15}$ mole/cm$^2$) over China; b) Time series of monthly mean NO$_2$ (blue), seasonal mean (orange for fall, red for summer, green for spring, purple for winter), and annual mean (black) NO$_2$ over the whole China; Horizontal distribution of year-to-year changes of NO$_2$ in (c) 2014-2015; (d) 2015-2016; (e) 2016-2017 and (f)2017- 2018.*

To investigate the latest changes of NO2 in different areas across China, we show a map of year-to-year change of NO2 from 2014 to 2018 (Figure 1c,1d,1e,and 1f). As expected, from 2014 to 2015, in all provinces across China, the change is negative or neglectable (within 0.1 unit, 1 unit= 10$^{15}$ mole cm$^{-2}$). From 2015 to 2016, the changes in most provinces are still negative, but the absolute values become smaller. From 2016 to 2017, most provinces in eastern China show strong positive (>0.5 unit) changes and most provinces in central and southern China also show positive changes. However, the changes in Beijing-Tianjing-Hebei are still negative. From 2017 to 2018, more provinces across the country show positive changes of NO2, particularly, in the northeastern China.

The histograms of year-to-year NO$_2$ changes for 31 provinces in mainland China sorted from large to small are shown in Figure 2. In 2015, only 2 of them (6.5%) show positive changes of NO2 from the previous year. However, the number increased to be 11(35.5%) , 23(74.2%) and 28(90.3%) in 2016, 2017 and 2018 respectively. The three provinces in 2017 with the largest increase of NO$_2$ are Jiangsu, Anhui and Zhejiang. In 2018, they are Tianjin, Beijing and Hebei. Time series of monthly mean NO$_2$ at 31 provinces are studied. Nine of them with the highest mean NO2 in Figure 1a are shown in Figure S1 (supplementary material).

The changes of $NO_2$ after 2016 in each province are also evident by looking at the year with maximum and minimal annual mean NO2 (refer to Figure S2 in the supplementary material). During 2015 to 2018, the NO2 in 2018 is the highest in all provinces except Shandong (2015), Yunnan (2016), Xizang (2017) and Anhui (2017).

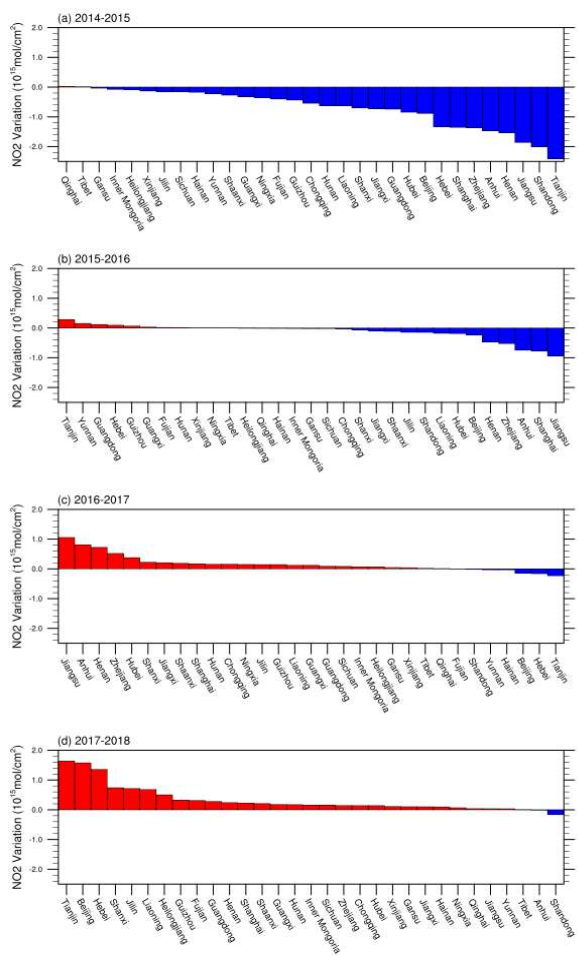

*Figure 2. Histograms of year-to-year change of NO2 column density in 31 provinces in mainland China sorted from large to small in (a) 2014-2015; (b)2015-2016; (c)2016-2017; and (d) 2017-2018.*

The principle of Aura OMI retrieving $NO_2$ column density is based on the selective absorption of solar radiation. However, the existence of cloud can significantly affect the retrieval and cause retrieving error. In the above studies, only satellite samples with cloud fraction less than 30% are used as in Liu et al 2016. To check if the selected threshold of "30%" will significantly affect the conclusion, the time series of annual mean NO2 across China with constrains of cloud fraction less than 10%, 20% and 30% are compared in Figure 3. It was found that, although changing the threshold of cloud fraction will lead to the changes (decreasing with cloud fractions threshold) of the amplitude of $NO_2$, the trends of increasing during 2005-2011 and the decline during 2011-2016 are not affected. Also, the increasing trend of $NO_2$ from 2016 to 2019 is constantly represented by the time series with different data filtering criteria (i.e the threshold of cloud fraction).

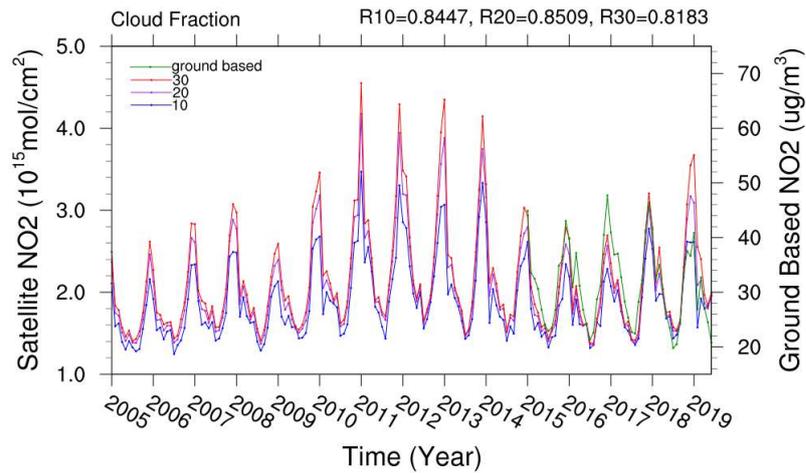

*Figure 3 Time series of monthly mean and national mean NO2 column density calculated using cloud-free samples with threshold of 30% (red), 20% (purple) and 10% (blue) from 2005 Jan to 2019 Jun. Overlapped (the green curve) is the associated ground-based measurements of NO2 concentration since 2015 Jan.*

Another error source comes from the hardware issue. It was found since 2007 Jun, there are some systematically questionable signals called "row anomalies" appeared in OMI observations. The retrieval due to row anomalies was already flagged in OMI NO2 product and was excluded from this study. But we still found some suspicious stripes appeared in the annual mean NO2 (refer to Figure S3a). This is due to the new development of row anomalies at the tracks 44, 45 and 46 (Figure S3c). Therefore, we further excluded those samples from the three tracks with $NO_2$ column density less than -$2\times10^{15}$ mole/cm². Those abnormal stripes in the annual mean NO2 were successfully removed (Figure S3b). The time series of monthly mean and national mean NO2 with and without this treatment of new row anomalies are shown in Figure S3d. It was seen that both of them constantly show the increasing trend of NO2 after 2016, particularly in winter and fall.

The time series of ground-based measurements of NO2 concentration from 1327 sites (the green curve in Figure 3) during 2015 to 2019 show strong positive temporal correlations with satellite column density (the correlation coefficients are 0.82, 0.85 and 0.84 for samples with cloud fraction less than 30, 20 and 10% ). However, the trend of annual mean and national mean NO2 from ground-based measurements is not as clear as that from satellite.

We further compared the year-to-year changes of NO2 in winter time from 2017 to 2019 (Figure 4) between satellite and ground observations. Both of them agree on a significant increase of NO2 in 2018 winter time in most areas in eastern and southern China, and in northern and northeastern China in 2019 DJF.

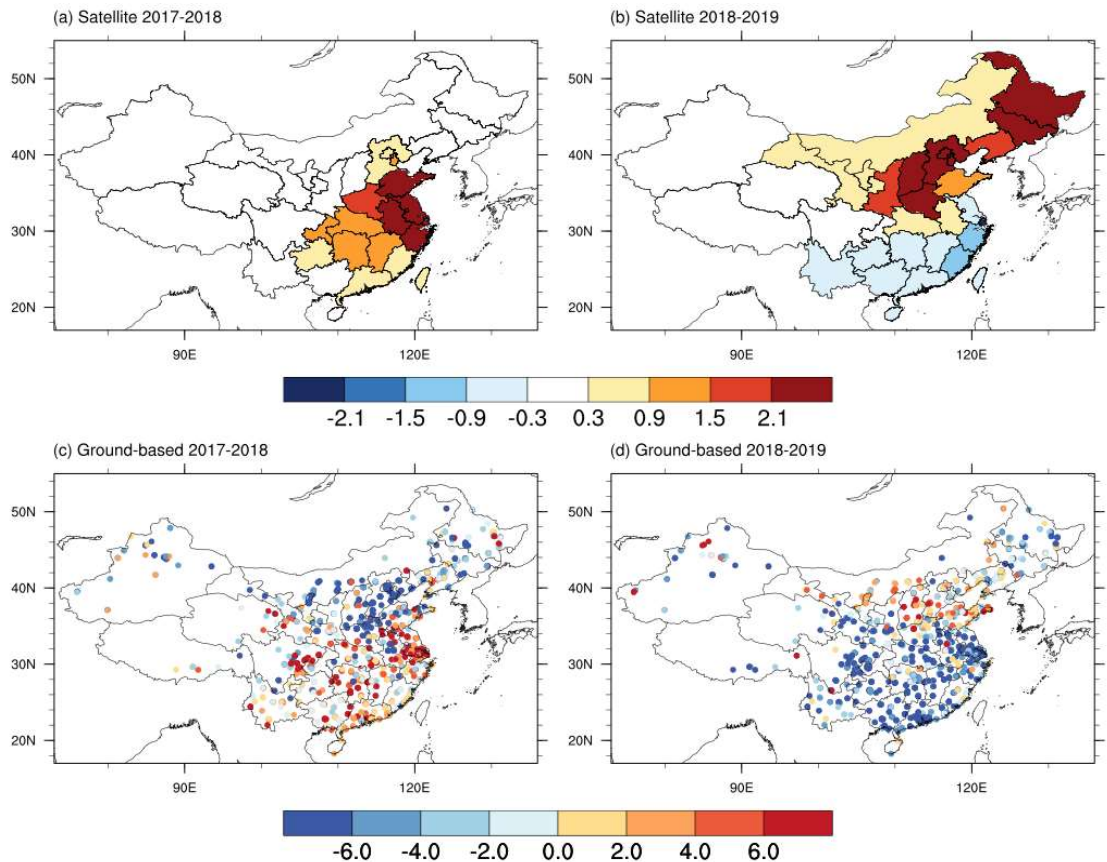

*Figure 4 The year-to-year changes of NO2 in winter time (DJF) derived from satellite column density (10$^{15}$ mole cm$^{-2}$, a and b) and ground measured concentration (ug m$^{-3}$, c and d) from 1372 sites cross China in the years 2017 to 2018 (a and c) and 2018 to 2019 (b and d).*

*Discussion*

State-of-the-art satellite remote sensing technology provides us a revolutionary way to monitor the air pollution world widely. With the aid of long-term observations of trace gas from Aura OMI, the trends of air pollution in the last decades are investigated world widely in several hot spots such as United States, United Kingdom, China, Russia etc. Those results overcome some of the shortcomings of ground measurements and provide better evidences for understanding the chemical and physical processes of air pollution. More importantly, the observed trends greatly helped policy makers to take controlling actions and legislations against the emission.

Based on the nearly 15 years (2005.1 to 2019.6) satellite observations from Aura OMI, the decline trend of tropospheric NO$_2$ from 2011 to 2015 is slowing down and overturned to be increasing in most provinces across China since 2016. This is unlikely to be caused by cloud contaminations and instrument anomalies. And the measurements from 1327 ground sites show strong positive temporal correlations with satellite observations. The maps of year-to-year changes of NO2 in winter time from satellite and ground observations show fairly well consistence.

Previous studies confirmed that the temporal variations of NO2 across China are mainly contributed by emissions from power plants and vehicles (Liu et al 2016, 2017). Since the nation-wide emission inventory in China has not been updated since 2015, it is not feasible to pin down the real reason of the observed new change of trend of tropospheric NO2 as seen from space. Further investigations of the reasons and the associated controlling actions must be done immediately to prevent worsening situation and to protect the achievement of air pollution controlling from the Twelve Five-Year plan (2010-2015).

*Data and Method*

The retrieved NO2 column data from the Dutch OMI tropospheric NO2 (DOMINO) v2.0 product (Boersma et al 2011 , available at TEMIS, http://www.temis.nl) from Jan 2015 Jan to Jun 2019 were used in this study. The introduction of Aura OMI is referring to Levelt et al., 2018. Aura is a polar orbit satellite with crossing time at equator 1:40 PM. The swath width of OMI is 2600 km and the horizontal resolution at nadir is 13 × 24 km2. To eliminate the contamination from cloud, only samples with cloud fraction <30% and surface albedo < 0.3 were used. Further studies about the sensitivity to the threshold of cloud fraction are referred to in the text. Row anomaly issue was found since Jun 2007, most likely was caused by an obstruction in part of the OMI's aperture (http://www.knmi.nl/omi/research/product/rowanomaly-background.php). We exclude those samples flagged with row anomalies from this study.

In addition, ground measurements of NO2 concentration at 1372 stations released from China National Environmental Monitoring Centre were also used in this study. The data is available at the official Ministry of Ecology and Environment download platform (http://106.37.208.233:20035/).

*Acknowledgment*

This work was supported by the National Key Research and Development Program of China (Grant No. 2017YFC1501402), the National Natural Science Foundation of China NSFC (Grant No. 41830104，41675022，41375148), Belmont Forum and JPI-Climate Collaborative Research Action with NSFC (Grant No. 41661144007), the "Hundred Talents Program" of the Chinese Academy of Sciences, the Hefei Institute of Physical Science (Grant No. 2014FXZY007), and the Jiangsu Provincial 2011 Program (Collaborative Innovation Center of Climate Change).

**Supplementary Material**

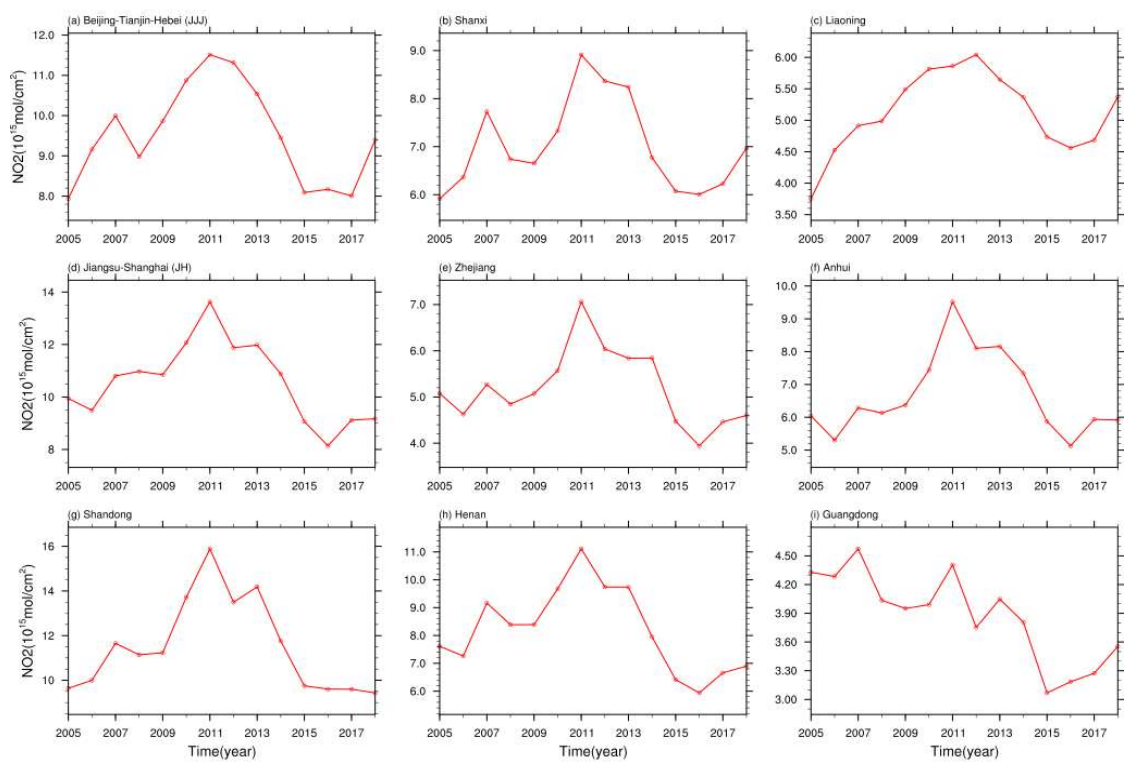

*Figure S1: Time series of monthly mean NO2 column density from Aura OMI in nine provinces in China.*

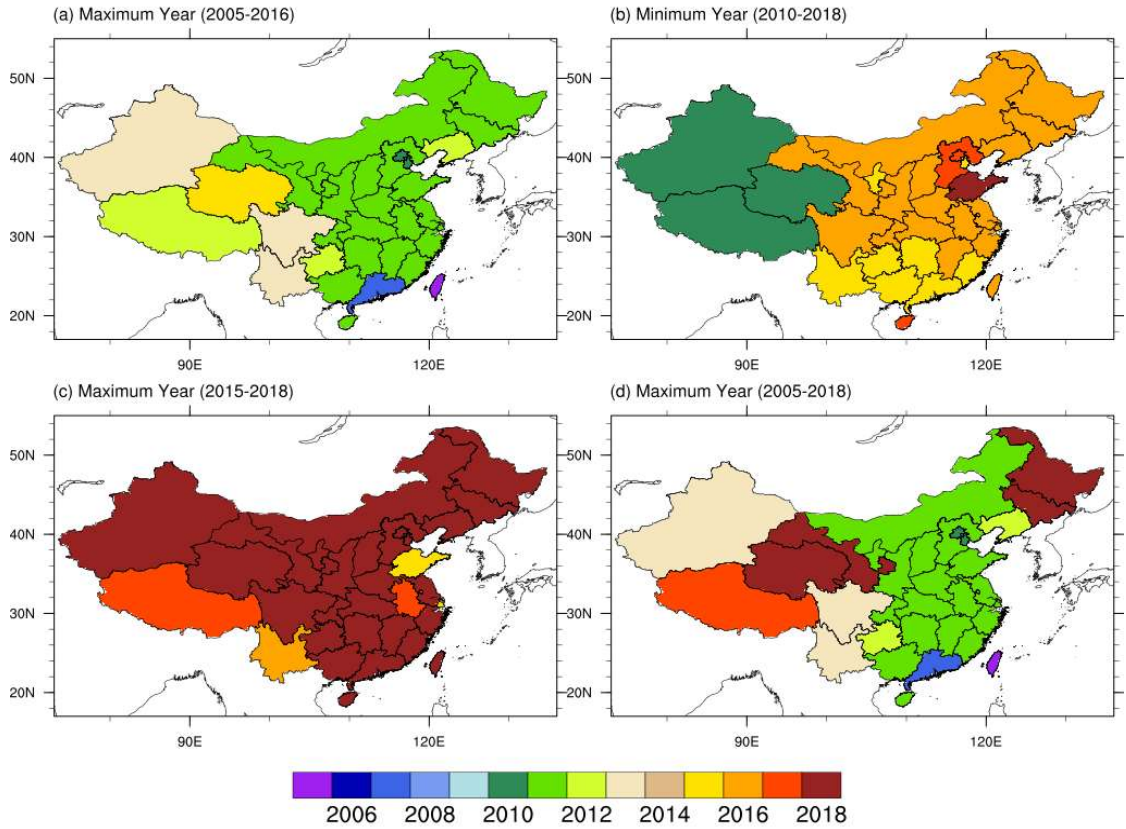

*Figure S2: The distribution of the years with maximum and minimum NO2 column density from Aura OMI in different time periods*

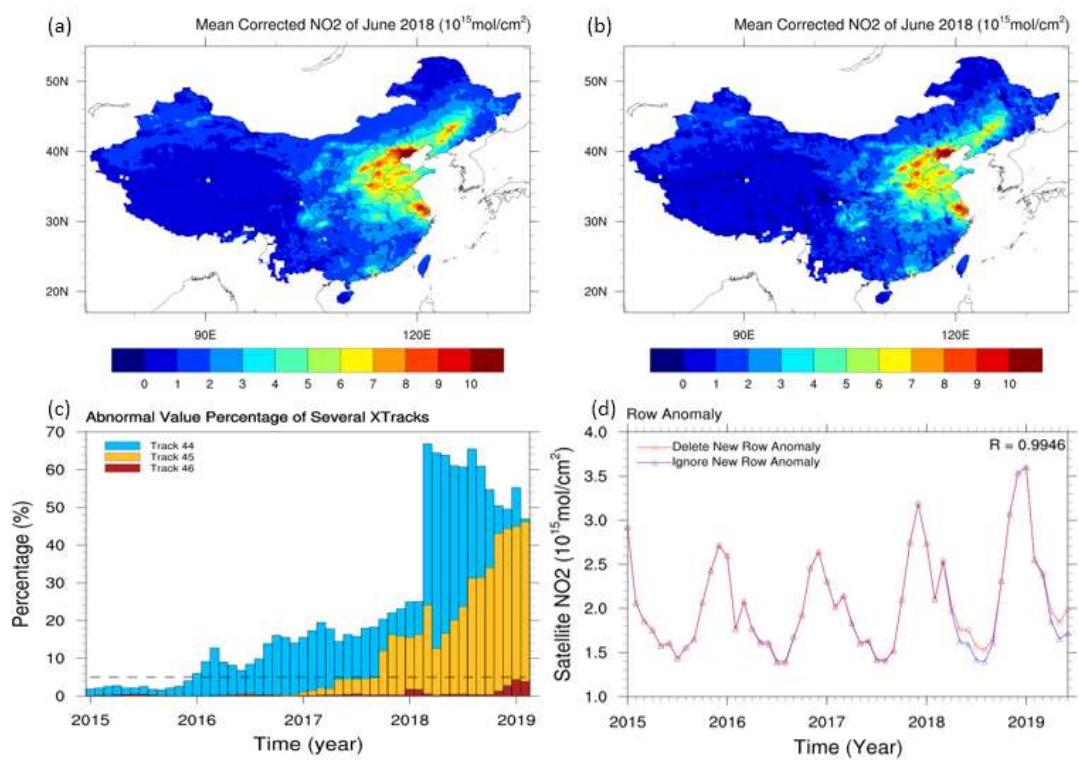

*Figure S3: (a) The mean NO2 column density (Aura OMI) in June 2018 after removing the new row anomalies; (b) same as (a), but without excluding the new row anomalies; (c) The monthly mean percentage of samples with the new row anomalies from 2015 Jan to 2019 Feb. (d) The time series of monthly mean and national mean NO2 column density (Aura OMI) from 2015 to 2019 with and without the treatment of new row anomaly.*